# When low- and high-energy electronic responses meet in cuprate superconductors

J. Chang[1], S. Pailhés[1], M. Shi[2], M. Månsson[3], T. Claesson[3], O. Tjernberg[3],
J. Voigt[4], V. Perez-Dieste[2], L. Patthey[2], N. Momono[5], M. Oda[5], M. Ido[5],
A. Schnyder[6], C. Mudry[6] and J. Mesot[1]*

[1] Laboratory for Neutron Scattering, ETH Zurich and PSI Villigen, CH-5232 Villigen PSI, Switzerland
[2] Swiss Light Source, Paul Scherrer Institute, CH-5232 Villigen PSI, Switzerland
[3] Materials Physics, Royal Institute of Technology KTH, S-164 40 Kista, Sweden
[4] Institute for solid state research, Forschungszentrum Jülich, 52425 Jülich, Germany
[5] Department of Physics, Hokkaido University - Sapporo 060-0810, Japan
[6] Condensed Matter Theory Group, CH-5232 Villigen PSI, Switzerland

* To whom correspondence should be addressed; E-mail: joel.mesot@psi.ch

**The existence of coherent quasiparticles near the Fermi energy in the low temperature state of high-temperature superconductors has been well established by angle-resolved photoemission spectroscopy (ARPES). This technique directly probes the momentum-resolved electronic excitation spectrum of the $CuO_2$ planes. We present a study of close to optimally doped $La_{1.83}Sr_{0.17}CuO_4$ in the superconducting state and report an abrupt change in the quasiparticle spectral function, as we follow the dispersion of the ARPES signal from the Fermi energy up to 0.6 eV. The interruption in the quasiparticle dispersion separates coherent quasiparticle peaks at low energies from broad incoherent excitations at high energies. We find that the boundary between these low-energy and high-energy features exhibits a cosine-shaped momentum de-**

**pendence, reminiscent of the superconducting $d$-wave gap. Further intriguing similarities between characteristics of the incoherent excitations and quasi-particle properties (lifetime, Fermi arcs) suggest a close relation between the electronic response at high and low energies in cuprate superconductors.**

**Introduction** High-temperature superconductors (HTSC) represent a prototype of strongly correlated systems that defies a description in terms of (one-electron) band theory. In these materials, the electrons interact strongly enough for the single-particle picture of band theory calculations to loose its validity, but not enough for the electron to be completely localized (for recent reviews see Refs. (*1–6*)). Surprisingly enough, angle-resolved photoemission spectroscopy (ARPES), that directly probes the momentum-resolved electronic excitation spectrum, has shown the emergence of single quasiparticles (QPs) upon entering the superconducting state of the HTSC $Bi_2Sr_2CaCu_2O_8$ (Bi2212) (*2, 4*). It is believed that pairs of these QPs constitute the charge carriers of high-temperature superconductivity. However, contrary to conventional superconductors, superconductivity in the cuprates is not simply caused by pairing of these quasiparticles. Rather, it is the process itself by which these quasiparticles emerge that seems to be the key to understanding the underlying mechanism (*1*). Experimental exploration of the composite nature of these quasiparticles and the mechanism that leads to their formation is crucial to understand the mechanism of high-temperature superconductivity.

By extracting an electron from the sample with a well defined energy and momentum, ARPES offers a unique insight into the electronic excitation spectrum of high-temperature superconductors. There is a solid body of photoemission measurements for binding energies of the order of 0.2 eV (relative to the Fermi energy $E_F$) that reveals many important features of high-temperature superconductors over a wide range of doping concentrations (*2, 4*). One example is the so-called kink, an anomaly in the quasiparticle dispersion that has been attributed to interactions with phonons (*7, 8*) or magnetic (*9, 10*) modes. While the dispersion of QPs in momentum space and for low binding energies has been extensively studied, their formation process, that might reveal itself at higher binding energies, has remained elusive. Only recently, ARPES studies on Bi-based cuprates have been performed that have uncovered unusual features at higher binding energies (0.2-1.5 eV) (*11, 12*).

Here, we report ARPES measurements on close to optimally doped $La_{1.83}Sr_{0.17}CuO_4$ in the superconducting state. This HTSC has a relatively simple crystal structure with a single $CuO_2$ layer. As we follow the dispersion of the QPs from $E_F$ up to 0.6 eV, we find an abrupt change in the spectral function characterized by a loss of intensity and by a considerable broadening of the peaks, thus revealing the complex composite nature of the low energy QPs. The interruption in the quasiparticle dispersion signals a transition from a quasiparticle behavior to an incoherent behavior. This boundary possesses a cosine-shape, very similar to the $d$-wave form of the superconducting gap. The close interplay between the spectral properties at low and high energies is underscored further by the fact that the high-energy incoherent excitations seem to have a memory for low-energy quasiparticles properties such as lifetime and Fermi arcs.

**Measurements** Our ARPES experiments were performed at the SIS beamline of the Swiss Light Source facility equipped with a SCIENTA SES2002 electron analyzer. We used 55 eV circularly polarized photons. The overall energy resolution was set to 35 meV and the angular resolution was 0.3°. The crystal ($T_c$ = 36 K), grown by the travelling solvent floating zone method (*13*), was aligned *ex situ* using Laue backscattering diffraction and then cleaved *in situ* at base temperature ($T$ = 15 K) under ultra high vacuum conditions. The quality of these crystals is excellent, as confirmed by the observation of a clear vortex lattice by means of neutron scattering (*14*). Data were recorded in swept mode from the second Brillouin zone, but are for convenience presented in the first zone. The detector was calibrated by recording spectra from polycrystalline copper on the sample holder.

**High-energy threshold $E_1$** We are going to present evidences for the existence of a characteristic energy scale $E_1$, measured relative to $E_F$, at which the composite nature of the low energy QPs becomes manifest. To this end, we plot in Figs. 1(a,b,c) the ARPES intensity as a function of binding energy and momentum along three cuts in the BZ (see inset in Fig. 1(c)). All three cuts show a high-intensity feature close to $E_F$. This feature has been extensively studied and interpreted as evidence for the existence of QPs (*2, 4*). A discontinuity in the slope of the QP dispersion that is well known for the cuprates and called a "kink" (green arrow in the inset of Figs. 1(a) and 1(b)) is observed at the binding energy $E_0 \approx 0.07$ eV. At the binding energy $E_1$ (white arrow in Figs. 1(a,b,c)), a less intense vertical feature emerges. Such a high-energy threshold $E_1$ was recently observed in the Bi2212 based family of cuprates (*11, 12*) and in the Mott insulator $Ca_2CuO_2Cl_2$ (*15*).

The excitation spectra can be analyzed via either momentum distribution curves (MDCs) or energy distribution curves (EDCs) by which is meant that the ARPES intensity is plotted either as a function of binding energy at fixed momentum or as a function of momentum at fixed binding energy (*2, 4*). EDCs and MDCs for the spectra in Fig. 1(a) are presented in Figs. 1(d) and (e), respectively. While peaks appear in the MDCs for all binding energies, the EDC peaks exist only between $E_F$ and $E_1$. Thus, fits to the Lorentzian line shapes of the MDCs provide the only way to extract the maximum intensities of the spectral function within the 0.6 eV large energy window. The position of the MDC peaks are depicted by a thin black line in the Figs. 1(a,b). Remarkably, in two cuts, Figs. 1(a,b), it is possible to identify *different* values for $E_1$ depicted by the white arrow, below which the thin black line becomes vertical. It is reasonable and indeed possible to fit the measured QP dispersion of the high-intensity feature using a tight-binding (TB) non-interacting model as explained in the caption of Fig. 1. However, since a QP is defined by a sharp peak in both EDCs and MDCs, it makes no sense to fit the thin black line for binding energies higher than $E_1$ on the basis of a single-particle picture. Inspection of Fig. 1(d) implies the existence of QPs with energies between $E_F$ and $E_1$ whereas the QP picture breaks down for binding energies higher than $E_1$.

Although the main purpose of this paper is to study the momentum dependence of $E_1$, we need to start with the Fermi surface as it is the reference for the low energy physics. The inset of Fig. 2(a) shows the Fermi surface, as determined by using MDC cuts at the fixed energy $E_F$, and eight selected momentum cuts for which we will gradually increase the binding energy from $E_F$ all the way to 0.6 eV. The cuts are numbered in ascending order as their intersections with the Fermi surface move from the nodal region to the anti-nodal region. The Fermi surface we report here is hole-like contrary to earlier claims (*16*). In Fig. 2(a) the relative MDC peak positions are shown as a function of binding energy (the color code is consistent with the inset). With increasing binding energy, the MDC peaks disperse until they reach $E_1$ denoted by black arrows. For binding energies higher than $E_1$ the MDC peaks are pinned in momentum. In some instances a colored arrow indicates a reentrance of dispersion. As the cuts approach the antinodal region the threshold value $E_1$ approaches $E_F$. In the case of cuts six, seven, and eight, two QP states are observed close to the Fermi level as indicated by two separate branches in Fig. 2(a). For cuts seven and eight these two branches merge at higher binding energy hereby defining $E_1$. An alternative characterization of the high-energy threshold $E_1$ can be done by analysing the half width at half maximum (HWHM) of the MDCs that we denote by $\Gamma_{\text{MDC}}(E)$ as a function of binding energy. Figs. 2(b1-8) show the energy dependence $\Gamma_{\text{MDC}}(E)$ for the eight cuts. The black arrows denote the threshold binding energy at which $\Gamma_{\text{MDC}}(E)$ becomes approximately constant at around 0.2 Å$^{-1}$, corresponding to a coherence length of a few lattice spacings. The arrows in Figs. 2(b1-8) correspond well to the black arrows in Fig. 2(a) for all cuts, except for cuts four and five for which they match the colored arrows in Fig. 2(a).

The white arrows in Figs. 1(a,b,c) and the black arrows in Figs. 2(a,b1-8) demonstrate that the high-energy threshold approaches the Fermi level when moving from the nodal to the antinodal region. Moreover, the locus of $E_1$ in reciprocal space traces a border between coherent and incoherent excitations that resembles the FS (see Fig. 2(b9)). We plot in Fig. 3 $E_1(\phi)$ as a function of the azimuthal FS angle $\phi$, defined in Fig. 2(b9). The red circles represent $E_1(\phi)$ as extracted from the onset of a vertical feature of the MDC peak positions (see the black arrows in Fig. 2(a)). The blue squares denote $E_1(\phi)$ as extracted from the onset of the saturation of the MDC HWHM (see the black arrows in Fig. 2(b1-8)). The dispersion $E_1(\phi)$ is well described by

$$E_1(\phi) = E_1(\pi/4) \times (1 - \cos 2\phi + \kappa). \tag{1}$$

For the doping level studied here, we found $E_1(\pi/4) = 0.43$ eV while we cannot resolve the value of $\kappa$, i.e., $\kappa E_1(\pi/4) < 0.035$ eV. The inset of Fig. 3 gives the doping dependence of $E_1(\pi/4)$ from the undoped (the Mott insulator $Ca_2CuO_2Cl_2$ sudied in Ref. (*15*)) to the strongly overdoped regime (SC Bi2212 studied in Ref. (*12*)). It is remarkable that the angular dependence of the $d$-wave gap, $\cos 2\phi$, a low-energy property of quasiparticles, enters in the dependence $E_1(\phi)$. Note that the decrease in energy of $E_1(\phi)$ between $\phi = \pi/4$ (nodal region) and $\phi = \pi/2$ (anti-nodal region) is considerable (0.4 eV). The boundary given by the data points in Fig. 3 delimits a coherent regime in which the excitations probed by ARPES are characterized by peaks in both EDC and MDC cuts from an incoherent regime in which excitations probed by ARPES are only characterized by broad MDC peaks. A consequence of the dispersion of $E_1$ is that, near the antinodal region, its energy becomes comparable to the energy of the kink, $E_0$. Where measurable, $E_0$ remains roughly constant along the FS as illustrated by the horizontal dashed line in Fig. 3.

The momentum dependence of the ARPES spectral weight at the fixed energies $E = 0, 0.1, 0.2, 0.3, 0.4$, and $0.5$ eV is shown in Figs. 4(a-f), respectively. As expected, at the lowest energy (Fig. 4(a)), the spectral weight is concentrated along the FS determined from the MDC peak positions (see the inset of Fig. 2(a)). At this low energy, sharp coherent QPs are observed at each FS point. The MDC widths are anisotropic and sharpest close to $0.4(\pi, \pi)$. As one moves up in binding energy, two changes can be noticed: first the locus of intensity along the $(\pi, \pi)$ directions moves toward $(\pi/4, \pi/4)$ and, second, the regions where QPs exist continuously shrink to single points along the diagonals (at $E_1(\pi/4)$), see Fig. 4(e), to eventually disappear completely for energies larger than $E_1(\pi/4))$ (Fig. 4(f)). Remarkably, at these high energies, the spectral function, although fully incoherent, remains strongly anisotropic. In Fig. 4 the dashed lines represent the crossover between coherent and incoherent regimes as defined in Fig. 3.

**Discussion** The paradigm of QPs has played a central role in condensed matter physics since its inception in the context of Fermi liquid (FL) theory. In the FL theory of the metallic or superconducting states, QPs evolve adiabatically from non-interacting electrons. On the FS, their life-time is infinite in a perfect crystal, i.e., QPs can be thought of as objects with all the attributes of an electron in vacuum except for a renormalized mass (and a gap for a SC). In a two-dimensional FL, the width of MDCs normal to the Fermi surface grow as the binding energy is increased relative to the Fermi energy with a corresponding increase in the width of the EDCs as the momenta move away from the Fermi surface towards the center of the BZ, the so-called $\Gamma$ point. Sufficiently far away from the Fermi surface, both EDCs and MDCs are featureless, i.e., once the inverse life times of QPs is comparable to their energies the notion of a QP is not applicable anymore. This expectation is confirmed by ARPES done on vicinal surface of Cu(111) (*17*).

ARPES has shown that the notion of QPs applies to LSCO deep in the SC state. QPs have emerged from what is believed to be a strongly correlated system. These QPs are objects with all the attributes of an electron (i.e., a charge and a spin-1/2) when probed with momenta and energies on the FS. The success of TB fits to the measured QP dispersion in the vicinity of the FS suggests that some effective one-band TB model supplemented with electron-electron or electron-phonon interactions can capture low energy features such as the kink seen by ARPES or the magnetic resonance and dispersing incommensurate peaks seen by inelastic neutron scattering (*18*). However, the existence of the energy scale $E_1$ proves that this notion of a QP must necessarily break down in an unexpected way compared to the breakdown of a QP in a conven-tional FL as it reveals a new characteristic length scale. This length scale, $\Gamma_{MDC}^{-1}(E_1)$ given by the MDC width at $E_1$, is of the order of a few lattice spacings. Beyond the energy scale $E_1$, this new length scale reveals that the low energy QPs have, in fact, an internal structure, as would be expected from an effective description of the cuprates in terms of a many-band Hubbard model (*19–23*). It is thus tempting to interpret $E_1$ as the energy scale for which a one-band Hubbard-like effective model breaks down in favor of a many-band Hubbard-like model. The characteristic length scale $\Gamma_{MDC}^{-1}(E_1)$ might then be related to the characteristic spatial extend of the O and Cu orbitals making up the Zhang-Rice singlet (*22*) in the one-band Hubbard-like effective low-energy and long-wave length model of HTSC. Although its seems unlikely that a one-band Hubbard-like model applies at energy scales of order 0.6 eV in LSCO, it is interesting to note that the pinning of the spectral weight along the vertical ridge from Fig. 1(a) resembles the pinning of the spectral weight for the one-hole spectral function in the $t-J$ model evidenced by Manousakis and interpreted by him as a transfer of spectral weight from the low energy QPs to higher-energy string-like excitations (*24*).

Exceptionally striking are the existing similarities between the low- and high-energy electronic responses. First, $E_1(\phi)$ is found to disperse in a way that closely tracks the $d$-wave SC gap. Notice, that as a consequence of this dispersion, the momentum points for which the high-energy anomaly is closest to the Fermi energy, correspond to points where the QP's lifetime is shortest (*25*). Second, by mapping the Brillouin zone as a function of increasing energy up to 0.6 eV it is found that the regions where QPs exist continuously shrink from a line (the Fermi surface at low energy), to a single point along the $(\pi, \pi)$ direction at $E_1(\phi = \pi/4) = 0.43$ eV (for this LSCO sample), to finally become fully incoherent. This behavior mimics that of the so-called Fermi-arcs as a function of temperature (*26*). Finally, notice that $E_1(\phi = \pi/4)$ has a doping dependence that resembles very much that of other energy scales, such as the gap maximum, the effective magnetic exchange $J$, or even the pseudogap in the underdoped regime

(see the inset of Fig. 3).

To conclude and in view of the similarities mentioned above, it is hard not to speculate that the same mechanism responsible for the incoherent high-energy features is also responsible for the emergence of coherent quasiparticles at much lower energies. From this point of view, the characterization of $E_1(\phi)$ puts severe constraints on the building of a microsocpic theory of high-$T_c$ superconductivity.

28. This work was supported by the Swiss National Science Foundation (through NCCR, MaNEP, and grant Nr 200020-105151), the Ministry of Education and Science of Japan and the Swedish Research Council. We thank C. Hess, M. Kropf, F. Dubi and P. Keller for technical support.

# Figures

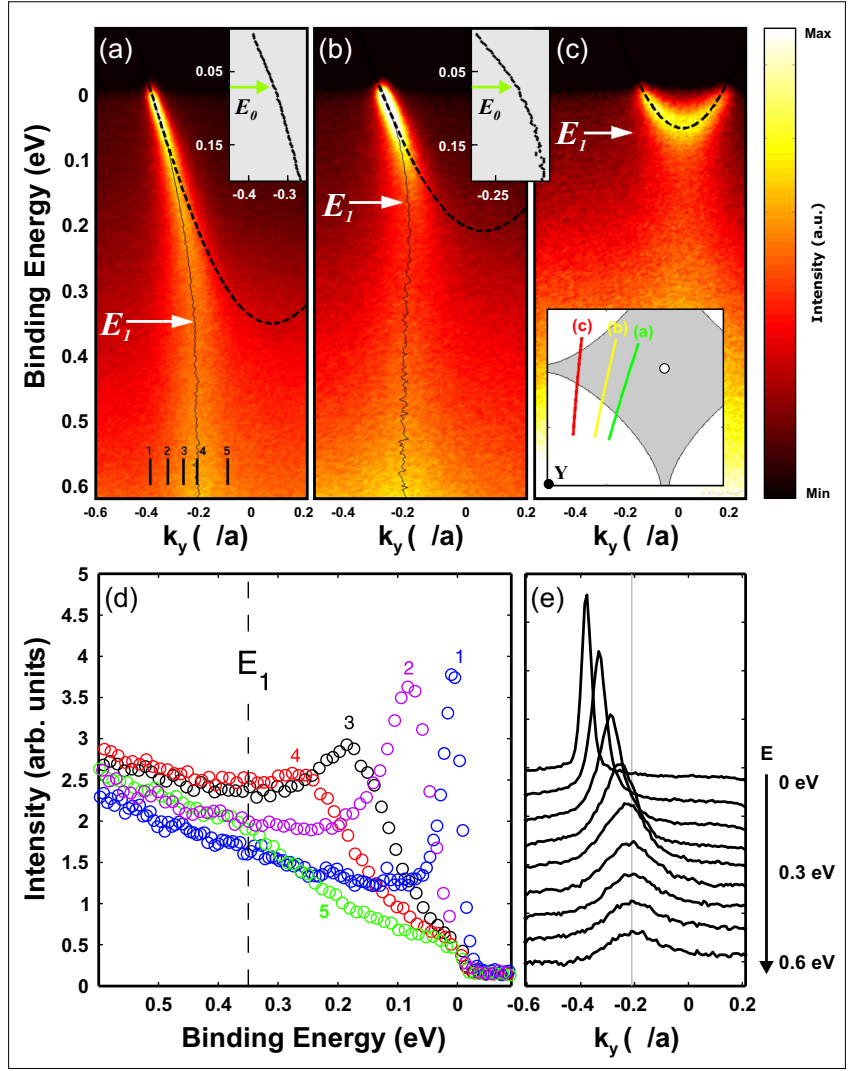

Figure 1: (a-c) Plots of ARPES intensity as a function of binding energy and momentum taken along the three cuts a, b, and c shown in the bottom-right inset. The intensity is given by a false color scale where white is the most intense. The thin black line represents the MDC peak positions. Zooms of the brightest part of the dispersion are shown in the top insets. The kink $E_0$ and the high-energy threshold $E_1$ are defined in the text. The dashed black line represents the tight-binding model (*27*) dispersion along each of the cuts in the Brillouin zone. The ratios $\mu/t = 0.84$, $t'/t = -0.144$, and $t''/t = 0.072$ are chosen so as to fit the measured FS. The band width, set by $t = 162.5$ meV, is determined by the measured Fermi velocity at the nodal point. (d) five EDCs of the spectra in (a) denoted by vertical lines in (a). (e) MDCs with binding energies between $E_F$ and 0.6 eV for the spectra shown in (a).

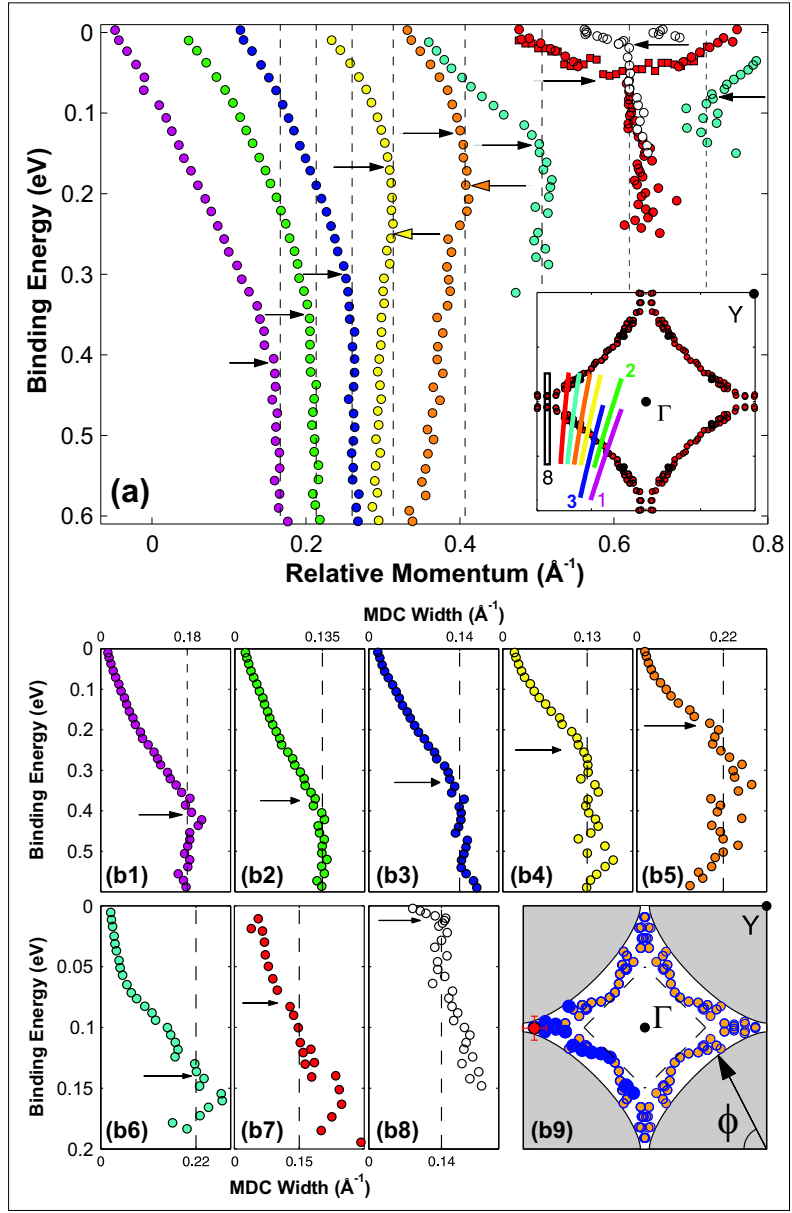

Figure 2: (a) MDC peak positions as a function of binding energy are shown for the eight different cuts from the inset. (The inset also shows the peak positions of MDCs at $E_F$, thereby mapping out the Fermi surface.) For cut seven, red squares correspond to the EDC peak positions. The horizontal scale indicates relative momentum position. (b1-b8) The binding energy dependence of $\Gamma_{\mathrm{MDC}}(E)$ for each of the eight cuts from the inset in (a). In figures (a) and (b), black arrows indicate the high-energy threshold. Colored arrows in (a) indicate reentrance of dispersion (see the text). The color code and the numbering are consistent with the inset in (a). (b9) Momentum position of $E_1$ as taken from (a) (symmetrized). Definition of angle $\phi$ along the Fermi surface in lower right corner.

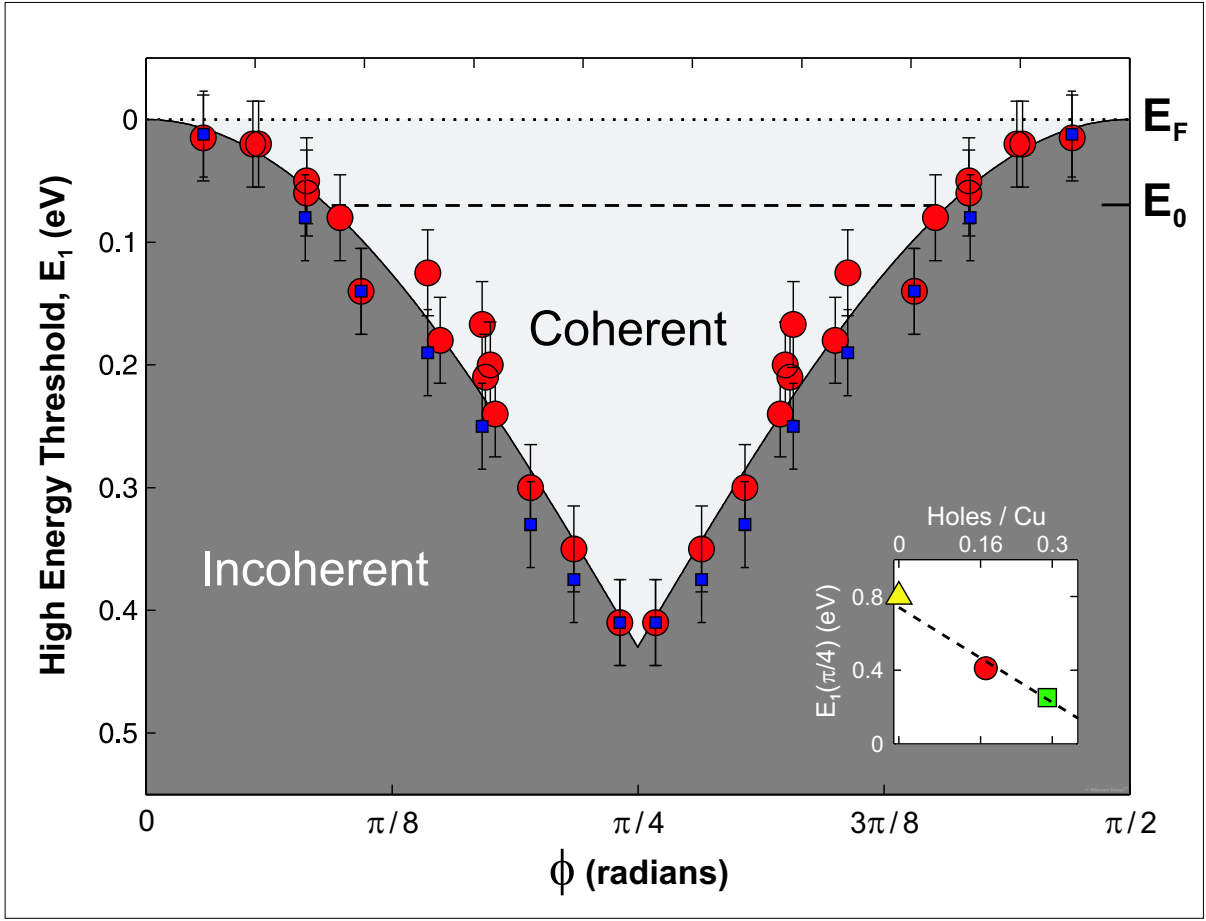

Figure 3: Dispersion of $E_1$ as a function of the angle $\phi$ defined in Fig. 2(b9). Red circles represent $E_1(\phi)$ as extracted from the onset of a vertical feature of the MDC peak positions (see the black arrows in Fig. 2(a)). Blue squares denote $E_1(\phi)$ as extracted from the onset of the saturation of the MDC HWHM (see the black arrows in Fig. 2(b1-8)). The inset shows $E_1(\phi = \pi/4)$ as a function of doping, yellow triangle from (*15*), red circle is this work and green square is from (*12*).

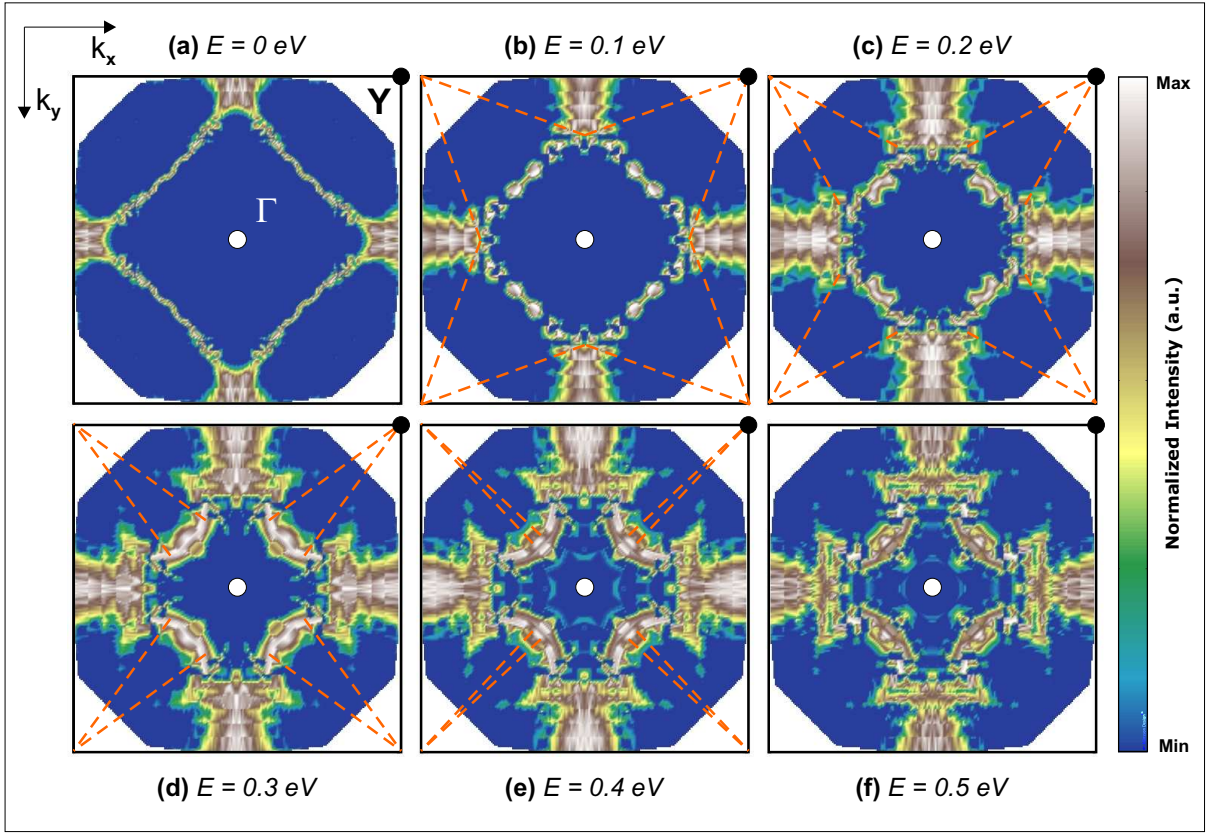

Figure 4: (a-f) Constant energy maps at binding energies $E = 0, 0.1, 0.2, 0.3, 0.4$, and $0.5$ eV. All maps are composed of background subtracted MDCs and the intensities have been normalized such that $I_{max} = 1$ for each MDC cut. Dashed lines represent the crossover between coherent (QP) and incoherent regimes as defined in Fig. 3.

17